\documentclass[10pt,twocolumn,letterpaper]{article}

\usepackage{cvpr}              

\usepackage{stfloats}
\usepackage{comment}
\usepackage[accsupp]{axessibility}

\newcommand{\FIGPATH}{figures_CVPR}

\newcommand{\widthframeworktwo}{0.46}

\newcommand{\widthdrag}{0.45}
\newcommand{\widthabl}{0.45}

\usepackage{soul, xcolor}

\newcommand\blfootnote[1]{%
  \begingroup
  \renewcommand\thefootnote{}\footnote{#1}%
  \addtocounter{footnote}{-1}%
  \endgroup
}

\definecolor{cvprblue}{rgb}{0.21,0.49,0.74}
\usepackage[pagebackref,breaklinks,colorlinks,allcolors=cvprblue]{hyperref}

\def\paperID{####} 
\def\confName{CVPR}
\def\confYear{2026}

\title{CraftMesh: High-Fidelity Generative Mesh Manipulation \\via Poisson Seamless Fusion}

\author{James Jincheng Hu$^1$ \quad Yuxiao Wu$^{2}$ \quad Youcheng Cai$^{2,*}$ \quad Ligang Liu$^{2}$\\
$^1$ Hefei Thomas School \quad $^2$ University of Science and Technology of China\\
{\tt\scriptsize contact@jameshu.org \quad wuyx2020@mail.ustc.edu.cn \quad caiyoucheng@ustc.edu.cn \quad lgliu@ustc.edu.cn}
}

\begin{document}

\twocolumn[{%
\renewcommand\twocolumn[1][]{#1}%
\maketitle
\centering
\vspace{-1.5em}
\includegraphics[width=1\linewidth]{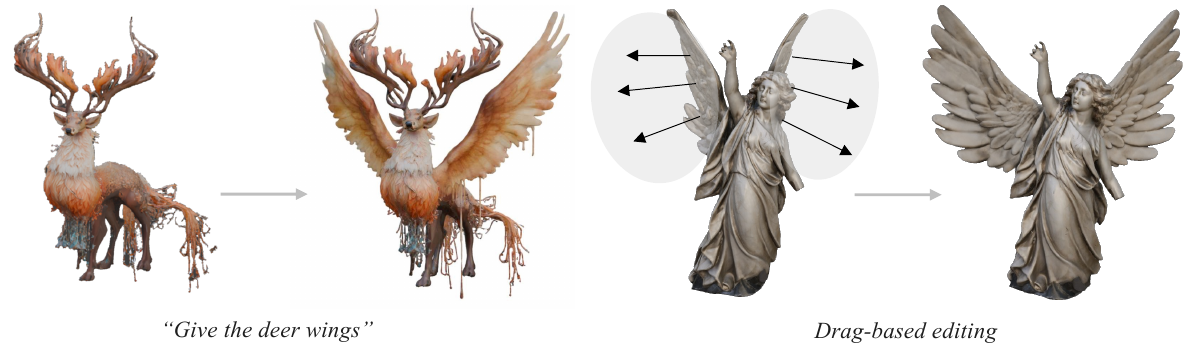}  
\captionof{figure}{Mesh editing results produced by CraftMesh. CraftMesh is a versatile 3D mesh editing framework that enables users to perform text-based and drag-based editing, and delivers high-quality outputs even in challenging editing scenarios.}
\label{fig:intro}

}]

\thispagestyle{empty} 

\begin{center}
  \large\bfseries Abstract
\end{center}
\textit{
Controllable, high-fidelity mesh editing remains a significant challenge in the domain of 3D content creation. Existing generative methods often struggle with complex geometries and fail to preserve fine-scale details. We propose CraftMesh, a novel framework for high-fidelity generative mesh manipulation based on Poisson Seamless Fusion. We decompose mesh editing into a pipeline that leverages the strengths of 2D image editing and 3D generation models: we first edit a 2D reference image, then generate a 3D mesh corresponding to the edited region, and fuse it seamlessly into the original mesh through a Geometry and Texture Fusion method. We introduce two core techniques: Poisson Geometric Fusion, which utilizes a hybrid SDF/Mesh representation with normal blending to achieve harmonious geometric integration, and Poisson Texture Harmonization for visually consistent texture blending. We demonstrate state-of-the-art structural consistency, geometric fidelity, and texture quality in challenging editing scenarios. 
}\\
Project page: \url{https://jameshu.org/CraftMesh}

\blfootnote{$^{*}$ Corresponding author.}
\section{Introduction}
In recent years, the rapid advancement of 3D generative models~\cite{lai2025hunyuan3d, xu2024instantmesh, li2024craftsman3d, siddiqui2024meta} has allowed the generation of high-quality 3D content from text prompts or images. These advances have substantially accelerated downstream applications in video games, augmented and virtual reality (AR/VR)~\cite{thi2025augmented}, robotics~\cite{liang2025diffusion}, and digital manufacturing~\cite{li2024generative}. Despite these notable achievements in 3D generation, the challenge of controllable 3D editing remains an open problem. Current 3D generation frameworks are designed to reconstruct 3D models from 2D images but provide limited flexibility for localized modifications.

Recent research has been done on neural field~\cite{mildenhall2021nerf} editing, encompassing appearance-guided and text- or image-driven approaches~\cite{haque2023instruct, wang2024gaussianeditor, zhuang2024tip}. These methods remain restricted to appearance-level modifications and cannot inherently support geometric manipulations of explicit surface meshes. In contrast to the rapidly expanding body of work on neural field editing, mesh-based generative editing has received considerably less attention, despite the fact that meshes remain the most widely adopted representation in professional 3D content creation pipelines. In practical design workflows, artists and engineers iteratively refine meshes with precise part-level control to satisfy both aesthetic and functional requirements while avoiding unintended alterations to unrelated geometry. This demand underscores the necessity for editing methods that enable fine-grained controllability while faithfully preserving the connectivity of the original model.

Existing generative mesh editing methods can be broadly categorized into two principal paradigms: score distillation sampling (SDS)-based approaches and multi-view diffusion (MVD)-based approaches. SDS-based methods~\cite{li2024focaldreamer, barda2024magicclay} optimize geometry using SDS losses and better retain the original structure, but they lack multi-view consistency, often producing oversimplified or distorted results. MVD-based methods~\cite{chen2024generic, barda2025instant3dit, li2025cmd} synthesize multi-view edits followed by a reconstruction step, but fail to preserve the geometry and texture of the original model.

To overcome these limitations, we harness the capabilities of generative models by reframing editing tasks as generative processes. We introduce an \textbf{image editing–mesh generation–seamless fusion} pipeline that fully leverages the strengths of 2D models for image editing and 3D models for high-quality mesh generation. The framework edits the rendered image of the target region, generates high-quality 3D content consistent with the edited view, and finally integrates the generated geometry and texture into the original mesh. However, a central challenge is achieving seamless geometric and textural consistency between the generated mesh and the source model.

Classical Poisson Mesh Editing \cite{yu2004mesh} operates in the coordinate domain but suffers from two fundamental limitations. First, incompatible connectivity—when transferring gradients from another mesh, this requires an unrealistic one-to-one vertex correspondence. Second, incompatible gradients—gradients in the coordinate domain are discontinuous at boundaries. SeamlessNeRF \cite{gong2023seamlessnerf} and GaussianStitching \cite{gao2025towards} blend radiance fields via Poisson-based optimization but cannot be directly applied to meshes with explicit surfaces. Directly solving the Poisson equation in 3D volumetric space incurs a computational complexity of $O(n^3)$, making it computationally expensive.

To this end, we present a novel high-fidelity generative mesh manipulation framework, termed \textbf{CraftMesh}, enabling complex and controllable editing while preserving geometry and texture (see Fig.~\ref{fig:intro}). We employ a 2D image editing model to modify reference images derived from the original mesh, generate corresponding meshes and extract the edited regions as edited region meshes, which are then fused into the original mesh via our core method, \textbf{Geometry and Texture Fusion}, an SDF-domain Poisson fusion method. We adopt a hybrid SDF/Mesh representation, in which geometry and texture are optimized in implicit and explicit forms, respectively. Specifically, We propose a \textit{Poisson Geometry Blending} strategy that enables natural gradient transitions and seamless geometric blending between the edited and original meshes. We further propose a \textit{Poisson Texture Harmonization} strategy to enable seamless texture fusion between the edited and original meshes in texture space. Experimental results demonstrate the superiority of our approach in achieving high-fidelity mesh editing. Additionally, we conduct further experiments using a drag-based method for controlled image manipulation, demonstrating the versatility of our framework (see \cref{fig:drag}).

Our contributions are summarized as follows:
\begin{itemize}
    \item We introduce a framework that reformulates mesh editing as an \textbf{image editing–mesh generation–seamless fusion} pipeline integrating 2D and 3D generative models.
    \item We propose Geometry and Texture Fusion, an SDF-domain Poisson fusion method, in which we introduce \textbf{Poisson Geometry Blending} to enable seamless geometric integration and \textbf{Poisson Texture Harmonization} to achieve seamless texture blending across edited regions.
\end{itemize}

\section{Related Work}
\,\,\,\,\,\,\textbf{3D Generation Models.} Recent advances in 2D diffusion models \cite{rombach2022high, oppenlaender2022creativity, labs2025flux} have significantly accelerated modern 3D content creation.

\textit{SDS-based Approaches}. Score Distillation Sampling (SDS) bridges 2D diffusion priors and 3D optimization. DreamFusion \cite{poole2022dreamfusion} first optimized NeRF under text-to-image diffusion guidance, followed by Magic3D \cite{lin2023magic3d}, which introduced a two-stage low-to-high resolution refinement. Later, LucidDreamer \cite{liang2024luciddreamer} further improved stability and fidelity through interval score matching, whereas ProlificDreamer \cite{wang2023prolificdreamer} incorporated a variational SDS formulation to improve diversity and quality. These methods successfully bridge 2D diffusion priors and 3D optimization, although they frequently remain computationally intensive.

\textit{MVD-based Approaches.} Multi-view diffusion (MVD) enforces multi-view consistency during image synthesis to reconstruct 3D assets. SyncDreamer \cite{liu2023syncdreamer} and MVDream \cite{shi2023mvdream} exploit multi-view diffusion for geometrically coherent text-to-3D generation. Wonder3D \cite{long2024wonder3d} and One-2-3-45++ \cite{liu2024one} further extend this paradigm to single-image 3D generation. SV3D \cite{voleti2024sv3d} uses latent video diffusion. Instant3D \cite{li2023instant3d} achieves fast reconstructions from sparse views.

\textit{3D Native Generation Approaches.} More recently, researchers have shifted toward training generative models directly on large-scale 3D datasets, thereby overcoming the inherent limitations of 2D priors. Foundational resources such as Objaverse \cite{deitke2023objaverse}, Objaverse-XL \cite{objaverseXL}, and OmniObject3D \cite{wu2023omniobject3d} provide millions of diverse, well-annotated 3D objects, enabling scalable learning of both geometry and texture. Clay \cite{zhang2024clay} demonstrates controllable large-scale text-to-3D generation. 3DTopia-XL \cite{chen20253dtopia} scales primitive-based diffusion approaches, achieving improved generalization across diverse categories. Trellis \cite{xiang2025structured} proposes structured 3D latent representations that improve scalability and versatility, making generative models more efficient at capturing complex shapes. Hunyuan3D 2.5 \cite{lai2025hunyuan3d} achieves high-fidelity geometry and physical-based rendering (PBR) texture. These native 3D models mark a shift toward more direct, efficient, and realistic 3D generation.

\textbf{Generative Mesh Editing.} Most existing generative editing approaches mainly focus on implicit representations \cite{haque2023instruct, wang2024gaussianeditor, zhuang2024tip, cheng2023progressive3d, sabat2024nerf}. Although these methods achieve promising results, they are constrained by implicit representations and thus cannot be applied to mesh-level editing. In this paper, we focus on generative mesh editing, which can be broadly categorized into two paradigms: SDS-based editing and MVD-based editing.

\begin{figure*}[t]
    \centering
    \includegraphics[width=1\textwidth]{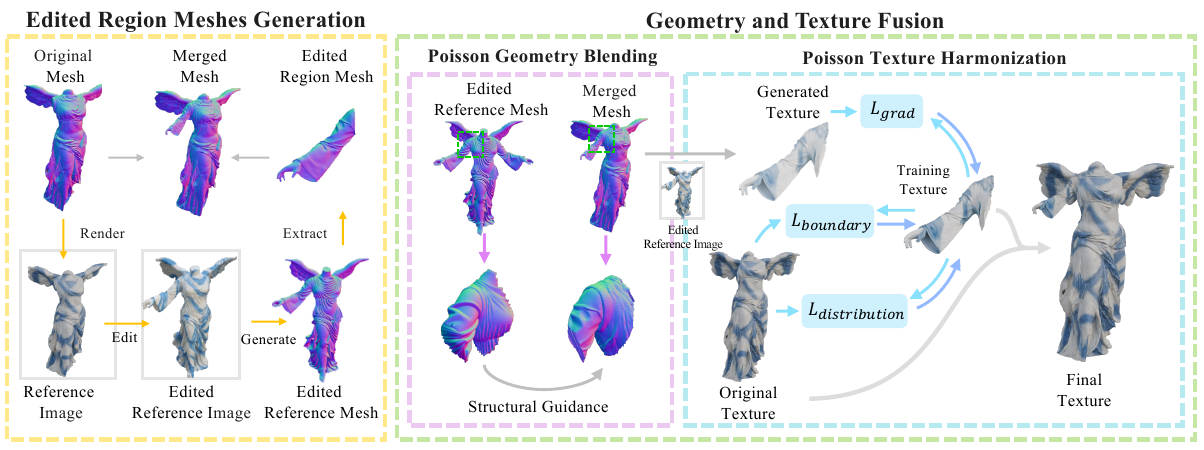}
    \caption{\label{fig:framework} An overview of the CraftMesh framework. Our framework follows an \textbf{image editing–mesh generation–seamless fusion} pipeline that fully leverages the strengths of 2D models for image editing and 3D models for high-quality mesh generation. First, \textbf{Edited Region Mesh Generation} produces meshes for the editing region. Then, \textbf{Poisson Geometry Blending} achieves natural geometric transitions through normal blending. Finally, \textbf{Poisson Texture Harmonization} performs texture fusion to seamlessly color the edited regions.}
\end{figure*}

\textit{SDS-based Editing.} SDS-based editing approaches extend the concept of Score Distillation Sampling (SDS) loss to editing tasks by guiding mesh optimization using pretrained diffusion priors. FocalDreamer \cite{li2024focaldreamer} introduces focal-fusion assembly for localized text-driven 3D editing, thereby enabling controllable, region-specific modifications. MagicClay \cite{barda2024magicclay} first optimizes an SDF with Score Distillation Sampling. Then, the changes are lifted to the editing mesh through vertex optimization. 

\textit{MVD-based Editing.} MVD-based editing methods utilize multi-view diffusion (MVD) to maintain consistency across different views, thereby enhancing fidelity. MVEdit \cite{chen2024generic} adapts generic 3D diffusion priors for controlled multi-view editing. CMD \cite{li2025cmd} proposes CondMV, which takes a target image and multi-view conditions to generate multi-view consistent edits. Instant3dit \cite{barda2025instant3dit} fine-tunes 2D diffusion models for multi-view consistent inpainting. MaskedLRM \cite{gao20243d} leverages large reconstruction models with masked conditioning for efficient mesh editing. 

However, these methods are unable to edit complex models or achieve high-fidelity mesh manipulation. In this paper, our method fully capitalizes on the strengths of 2D and 3D generative models. By employing a Poisson seamless fusion strategy, our approach merges generated edited region meshes with the original mesh, thereby achieving high-fidelity and structurally consistent mesh manipulation.

\textbf{Seamless Editing.} Pioneered by Poisson Image Editing \cite{perez2003poisson}, seamless editing is a classical topic in computer graphics and digital image processing. The primary goal is to achieve smooth and imperceptible transitions in images or textures, thus maintaining visual consistency \cite{agarwala2004interactive, kwatra2005texture, barnes2009patchmatch}. Poisson Mesh Editing \cite{yu2004mesh} applies the Poisson equation to mesh editing, enabling geometric merging through gradient-field manipulation. However, these methods solve the Poisson equation on the coordinate domain, which results in discontinuous gradients at the merging boundary that create artifacts. Our method solves the Poisson equation on the gradient domain.

Recently, SeamlessNeRF \cite{gong2023seamlessnerf} achieves seamless blending of neural radiance fields through a Poisson based optimization, focusing on radiance field merging. Gao et al.~\cite{gao2025towards} advances example-based 3D modeling by introducing 3D Gaussian stitching. While these works offer smooth merging in radiance fields or 3D Gaussians, explicit mesh geometry is not considered. In this paper, we consider both geometry and texture, ensuring seamless fusion between the edited region mesh and the original mesh.
\section{Method}

We propose CraftMesh, a high-fidelity generative mesh manipulation framework that integrates 2D image editing, 3D mesh generation, and Poisson-based fusion. \cref{fig:framework} illustrates the overall workflow. We first edit reference images using 2D image editing models to achieve user-intent-consistent modifications, followed by generating edited region meshes with 3D generative models. Next, we employ a Poisson Seamless Fusion strategy to integrate the edited-region mesh into the original mesh, ensuring geometric and texture consistency through Poisson Geometry Blending and Poisson Texture Harmonization.

\subsection{Edited Region Meshes Generation}
Recent image generation models have demonstrated remarkable performance in controllable image editing, producing semantically aligned and globally consistent results. Representative examples include FLUX Kontext~\cite{labs2025flux}, Qwen-Image \cite{yang2025qwenimage}, and Gemini 2.5 \cite{comanici2025gemini}, which can effectively preserve content structure while introducing new details. Compared with direct 3D editing, these 2D approaches are lightweight, controllable, and well-suited for performing high-quality object-centric image manipulation. On the other hand, recent progress in 3D mesh generation, such as Craftsman3d \cite{li2024craftsman3d} and Hunyuan3D \cite{lai2025hunyuan3d}, has enabled the production of 3D meshes with unprecedented structural fidelity, intricate details and textural realism. However, existing 3D mesh editing methods lag significantly behind. For instance, Instant3dit~\cite {barda2025instant3dit} fine-tunes multi-view diffusion models to regenerate 3D content, but often struggles with consistency. Similarly, FocalDreamer \cite{li2024focaldreamer} and MagicClay \cite{barda2024magicclay} are limited to basic tasks and frequently yield simple results in the edited region.

To bridge this gap, we leverage the complementary advantages of 2D image editing models and 3D mesh generation models. Specifically, we first generate \textbf{Edited Region Meshes}, which we later fuse with the original mesh (see \cref{sec:geo}, \cref{sec:tex}). The generation proceeds as follows: From the original mesh, we render the reference image, which is edited according to the user’s intent of mesh editing. Users can employ a variety of tools to perform the edit, including image editing models \cite{labs2025flux}, software suites, or other creative instruments. This flexibility in choosing the image-editing backend provides users with greater control and adaptability to achieve the desired outcome. Next, we lift the edited reference image to 3D using image-to-mesh generation models \cite{xiang2025structured, lai2025hunyuan3d, meshy2025}. From this edited reference mesh, we extract the components within the editing region to form the edited region mesh. Optionally, we can perform an additional image editing step on the edited reference image to extract an image of the editing region, which is used by image-to-mesh models to generate the edited region mesh with higher fidelity. 

\begin{figure}[t]
    \centering
    \includegraphics[width=\widthframeworktwo\textwidth]{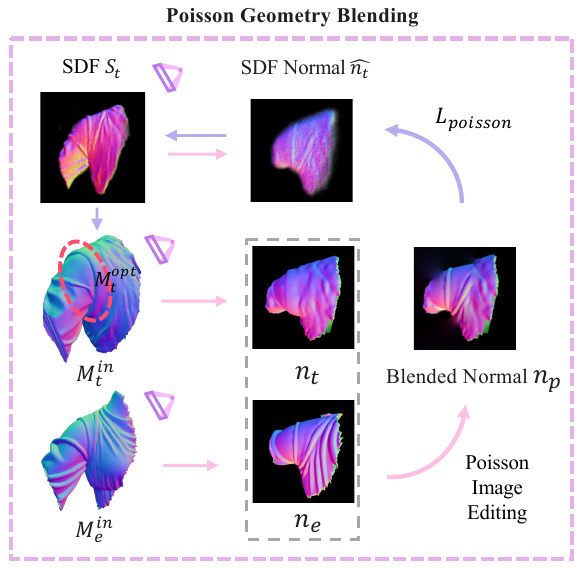}
    \caption{\label{fig:framework_2} Details of Poisson Geometry Blending. We employ a hybrid SDF/Mesh representation, which is optimized to be harmonious using blended normals.}
\end{figure}

Our core idea is to seamlessly fuse the edited region mesh into the original mesh while using the edited reference mesh as geometric guidance. CraftMesh offers: (1) control of mesh editing through image editing, meaning no requirement of manually specifying editing locations in 3D, unlike FocalDreamer~\cite{li2024focaldreamer}, MagicClay~\cite{barda2024magicclay}, and Instant3dit~\cite{barda2025instant3dit}, making editing more controllable and user-friendly; (2) effective integration of 2D editing models with 3D mesh generation, achieving high-quality edited region meshes with structural harmony.

\subsection{Geometry and Texture Fusion}

We apply geometry and texture fusion on a hybrid SDF/Mesh representation \cite{barda2024magicclay}. For geometry, indirectly optimizing an SDF rather than directly optimizing mesh vertices yields more stable convergence and improved robustness to noise. For texture, we represent color through an implicit neural color field. Neural SDF fields provide several advantages over discrete meshes, including robustness to noise, improved convergence, differentiable rendering, analytical gradients, and inherent continuity. We leverage these properties to perform Poisson seamless fusion in the SDF domain, enabling simultaneous and coherent fusion of geometry and texture.

\subsubsection{Poisson Geometry Blending}
\label{sec:geo}

Based on the Poisson equation, we propose a Poisson Geometry Blending strategy to achieve seamless integration of the edited region into the original mesh while leveraging the edited reference mesh as structural guidance. We ensure that the final mesh inherits the seamless transition and natural details of the reference mesh while retaining the fidelity of the original mesh and edited region mesh.

\cref{fig:framework_2} gives an overview of the workflow. We first apply a mesh Boolean operation \cite{cherchi2022interactive} between the original mesh and the edited region mesh to generate an initial merged mesh. We align the original mesh with the edited region mesh according to \cite{zhu2025imaginariumvisionguidedhighquality3d}. We then adopt a hybrid SDF/Mesh representation, which enables flexible refinement of mesh geometry by optimizing vertex positions, splitting triangles and collapsing edges. The refinement is guided by normal maps rendered from both the edited reference mesh and the edited region mesh, which are blended using a Poisson-based approach. This geometry blending strategy allows the edited region mesh to be naturally incorporated into the original mesh with harmonious boundary transitions.

\textbf{Intersection Region Extraction.}
Given the original mesh $M_o$ and the edited region mesh $M_r$, we first apply a mesh Boolean operation to obtain a merged mesh $M_t$. For insertion tasks, we use mesh Boolean union, and for deletion tasks, we use mesh Boolean difference. Geometric discontinuities occur at the transition boundary, so we extract and later refine it using a hybrid SDF/Mesh representation. The extraction process is as follows:

The Boolean operation produces a set of vertices $V_{in}$ at the intersection between $M_o$ and $M_r$. We align the edited reference mesh $M_e$ with $M_t$. The corresponding intersection regions of $M_t$ and $M_e$ can be defined as:
\begin{align}
&M_t^{in}=\left\{v \in M_t \mid \min_{u \in V_{in}}\|u-v\|_2 < \epsilon_0 \right\},\\
& M_e^{in}=\left\{v \in M_e \mid \min_{u \in V_{in}}\|u-v\|_2 < \epsilon_0 \right\},
\end{align}
where $\epsilon_0$ controls the extent of the intersection. This focuses attention on the harsh transition. We define a smaller subset of  $M_t^{in}$ as its optimization region:
\begin{equation}
M_t^{opt}=\left\{v \in M_t^{in} \mid \min_{u \in V_{in}}\|u-v\|_2 < \epsilon_1\right\}, \quad \epsilon_1 < \epsilon_0.
\end{equation}

\textbf{Poisson Normal Blending Guidance.}
We train the mesh $M_t^{in}$ indirectly by optimizing a neural SDF $S_t$ it is bound to, changes in the SDF will be propagated to the mesh through vertex optimization by refining vertex positions, splitting triangles and collapsing edges. Vertices not in the optimization region $M_t^{opt}$ are frozen during optimization. Indirectly optimizing an SDF instead of direct mesh optimization offers stable convergence and robustness against noise. Compared to voxel-based methods like DMTet \cite{munkberg2022extracting}, vertex optimization achieves precise and natural topology.

During optimization, we render multi-view supervision images: (1) a normal map of $M_t^{in}$, denoted $n_t$; (2) a binary mask of $M_t^{opt}$, denoted $mask^{opt}$; (3) a normal map rendered from the SDF $S_t$, denoted $\hat{n}_t$; (4) a normal map of $M_e^{in}$, denoted $n_e$. 
To enforce consistency, we apply the classical Poisson Image Editing (PIE) algorithm \cite{perez2003poisson} $\Gamma(\cdot)$ to blend $n_t$ and $n_e$ under $mask^{opt}$:
\begin{equation}
n_p = \Gamma(n_t, n_e, mask^{opt}).
\end{equation}
The blended normal map $n_p$ preserves fine-grained details from $n_e$ inside the mask while achieving a seamless transition to $n_t$ along the mask boundary. Solving the Poisson equation on 2D images reduces computational complexity to $O(kn^2)$, offering significant efficiency over 3D volumetric methods ($O(n^3)$) while maintaining comparable fidelity.

We then minimize the discrepancy between the rendered normal $\hat{n}_t$ and the blended normal $n_p$:
\begin{equation}
\mathcal{L}_{\text{poisson}} = \sum_i\|\hat{n}_t^i - n_p^i\|_F^2,
\end{equation}
where $\|\cdot\|_F$ denotes the Frobenius norm and $i$ indexes different camera viewpoints. Although the blended normal maps $n_p^i$ are not strictly multi-view consistent, the implicit SDF effectively resolves inconsistencies and learns a coherent transition geometry. Following MagicClay \cite{barda2024magicclay}, we further incorporate additional regularization terms, such as a smoothness loss $\mathcal{L}_{\text{smooth}}$ and an Eikonal loss $\mathcal{L}_{\text{eik}}$, to improve geometric fidelity and enforce implicit surface constraints. The loss for geometry blending is formulated as:
\begin{equation}
\mathcal{L}_{\text{geo}} = \mathcal{L}_{\text{poisson}} + \lambda_1 \mathcal{L}_{\text{smooth}} + \lambda_2 \mathcal{L}_{\text{eik}},
\end{equation}
where $\lambda_1$ and $\lambda_2$ are parameters. 

\subsubsection{Poisson Texture Harmonization}
\label{sec:tex}
The newly synthesized regions of the merged mesh $M_t$ lack texture information. A straightforward solution is to employ a texture generation model \cite{meshy2025} conditioned on the edited reference image for color generation; however, the resulting textures often exhibit noticeable shifted colors compared to the original mesh and discontinuities along boundaries, as shown in \cref{fig:framework}. Therefore, we adopt the following Poisson-based strategies to improve overall visual harmony.

\textbf{Gradient Propagation} 
Let $C_{new}$ denote the color field of $M_t^{new}$. 
Before optimization, we store a frozen copy $C_{new}^{ori}$ for gradient reference. 
Since gradients encode details, we retain fine texture patterns by enforcing consistency between the gradients of the current and original color fields:
\begin{equation}
\mathcal{L}_{\text{grad}} = 
\text{MSE}\!\left(
\sigma\!\left(\frac{\nabla C_{new}}{\gamma}\right),
\sigma\!\left(\frac{\nabla C_{new}^{ori}}{\gamma}\right)
\right),
\label{eq:Lgrad}
\end{equation}
where $\nabla$ denotes numerical gradients, $\sigma(\cdot)$ is the sigmoid function, and $\gamma$ is a gradient scaling constant. This term stabilizes and preserves high-frequency texture details.

\textbf{Smooth Transition Refinement.} 
To achieve seamless blending at intersection boundaries, we introduce a distance-weighted soft boundary loss:
\begin{equation}
\mathcal{L}_{\text{boundary}} =
\sum_{p_i^{new} \in M_t^{new}}
w_i \| C_{new}(p_i^{new}) - C_{pr}(p_i^{pr}) \|_2^2,
\label{eq:Lcolor}
\end{equation}
where $p_i^{pr}$ is the nearest point on $M_t^{pr}$ to $p_i^{new}$, and 
\begin{equation}
w_i = \left(1 - \frac{\delta}{\|p_i^{new} - p_i^{pr}\|_2}\right)^2
\end{equation}
attenuates the influence with distance. 
The parameter $\delta$ controls the effective boundary width. This boundary loss enforces smooth color transitions and prevents visible seams between fused regions.

\textbf{Distribution-Aware Color Alignment.} 
Theoretically, $\mathcal{L}_{\text{grad}}$ and $\mathcal{L}_{\text{boundary}}$ follow the formulation of Poisson Image Editing (PIE) \cite{perez2003poisson}. However, the repetitive patterns inherent in textures interfere with gradient-domain color propagation, causing failure to normalize colors (Fig.~\ref{fig:abl}e). Therefore, we propose an $\mathcal{L}_{\text{distribution}}$ loss that addresses this limitation by enforcing color distribution consistency between the generated and preserved regions.

Let $M_t^{new}$ denote the newly synthesized geometry and $M_t^{pr}$ the preserved geometry from the original mesh. The preserved mesh $M_t^{pr}$ inherits textures directly from the original, while $M_t^{new}$ is textured using a generative model \cite{meshy2025}. The predicted colors form a probability density distribution in RGB space, which is obtained using kernel density estimation:
\begin{equation}
    \rho(q) = \frac{1}{N} \sum_{i=1}^{N} \exp\!\left(-\frac{\| q - r_i \|^2}{2\sigma^2}\right),
    \label{eq:density}
\end{equation}
where $\{r_i\}_{i=1}^{N}$ are sampled mesh colors and $\sigma$ denotes the Gaussian kernel bandwidth.  

We denote the color distributions of $M_t^{new}$ and $M_t^{pr}$ as $\rho^{new}$ and $\rho^{pr}$, respectively. 
Distribution-aware alignment is achieved by minimizing their discrepancy:
\begin{equation}
\mathcal{L}_{\text{distribution}} = \frac{1}{N} \sum_{i=1}^{N} \|\rho^{new}(q_i)-\rho^{pr}(q_i)\|_2,
\label{eq:Ldensity}
\end{equation}
where $\{q_i\}_{i=1}^N$ are color samples from $M_t^{new}$.

The overall optimization objective for Poisson Texture Harmonization is:
\begin{equation}
\mathcal{L}_{\text{tex}} =
\mathcal{L}_{\text{distribution}} +
\theta_1 \mathcal{L}_{\text{grad}} +
\theta_2 \mathcal{L}_{\text{boundary}},
\label{eq:Ltex}
\end{equation}
where $\theta_1$ and $\theta_2$ balance the gradient and boundary consistency terms.
Unlike prior mesh editing pipelines that synthesize only albedo materials, our formulation directly extends to physically based rendering (PBR) materials.
\begin{figure*}[t]
    \centering
    \includegraphics[width=1\textwidth]{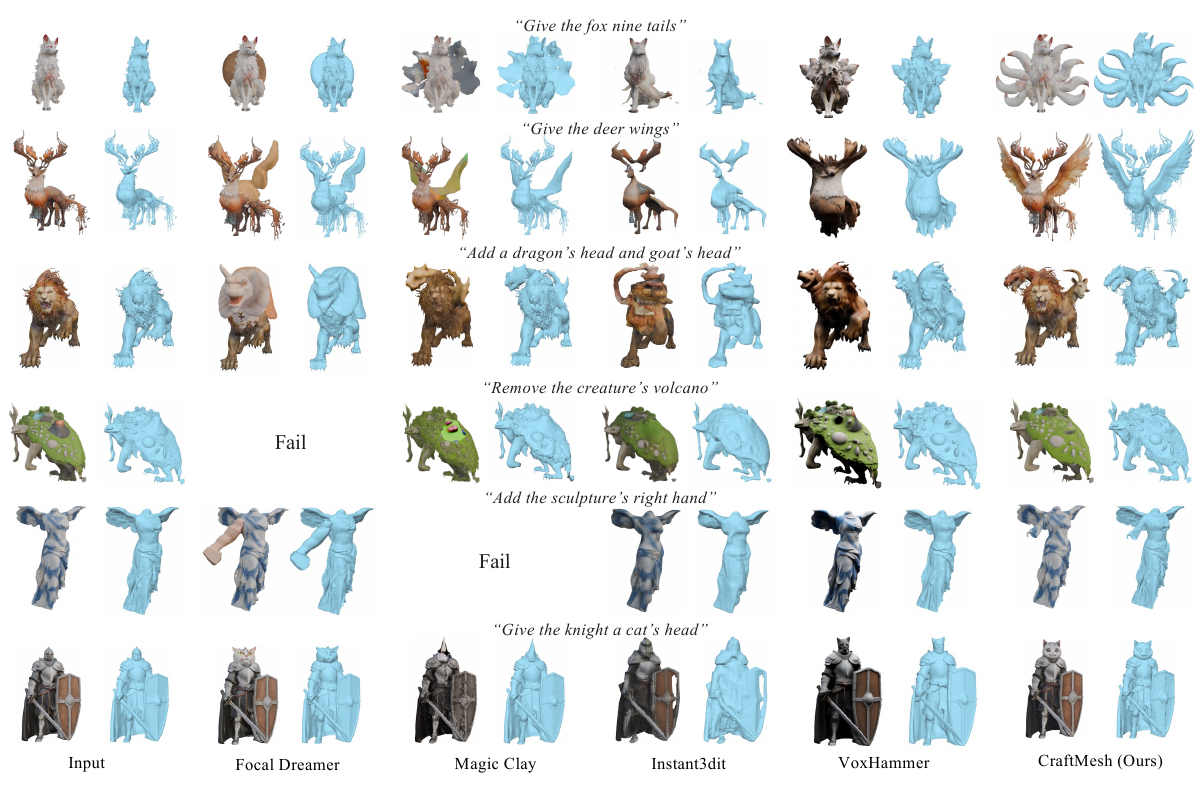}
    \caption{\label{fig:compare} Qualitative comparisons show that our method produces harmonious geometry structure, intricate local details, and high-fidelity colors, while other methods give simple and inconsistent results.}
\end{figure*}

\section{Experiments}

\subsection{Experiment Setup}
\,\,\,\,\,\, \textbf{Implementation.} We use FLUX Kontext \cite{labs2025flux} as the generative image-editing model and MeshyAI \cite{meshy2025} as the image-to-mesh generator. It is worth noting that our framework is model-agnostic with respect to these components. In the future, more powerful models can be readily incorporated when conducting experiments. We adopt MagicClay \cite{barda2024magicclay} as the backbone for both the hybrid SDF–mesh representation and the implicit neural color field. On a single 4090 GPU, Poisson Geometry Blending requires 5 minutes for 1000 iterations, while Poisson Texture Harmonization requires 1 minute for 2000 iterations.

\textbf{Dataset.} We construct an evaluation dataset comprising 100 diverse and complex 3D models, selecting from Objaverse-XL~\cite{objaverseXL}, Google Scanned Objects~\cite{downs2022googlescannedobjectshighquality} {and the internet. For each model, we apply 3 distinct editing prompts. We evaluate these meshes through a series of complex editing tasks, including insertion, deletion, and localized region editing, to comprehensively assess our method’s effectiveness in achieving global structural consistency and high-quality local details.

\textbf{Baselines.} We compare our method with recent state-of-the-art approaches for mesh editing, including SDS-based methods such as FocalDreamer~\cite{li2024focaldreamer} and MagicClay~\cite{barda2024magicclay}, the MVD-based method Instant3dit~\cite{barda2025instant3dit}, and the Latent-based method Voxhammer~\cite{li2025voxhammer}. All results were obtained using their official open-source implementations.

\begin{table}[t]
\centering
\resizebox{\columnwidth}{!}{
\begin{tabular}{l|cccc}
\toprule
Method & CLIP\textsubscript{sim} $\uparrow$ & CLIP\textsubscript{dir} $\uparrow$ & NIQE $\downarrow$ & NIMA $\uparrow$ \\
\midrule
FocalDreamer & 13.010 & 3.927 & 12.340 & 5.234 \\
MagicClay & 15.043 & 5.994 & 7.344 & 5.334 \\
Instant3dit & 14.108 & 4.326 & 7.390 & 5.288 \\
VoxHammer & 17.366 & 10.482 & 8.291 & 5.307 \\
Ours (MeshyAI) & \textbf{20.801} & \textbf{18.479} & \textbf{4.710} & \textbf{5.928} \\
\bottomrule
\end{tabular}
}
\caption{Quantitative comparison with baselines. Our method consistently achieves the best performance across all metrics, demonstrating better semantic consistency and visual quality compared to existing methods.}
\label{tab:compare_quant}
\end{table}

\subsection{Quantitative Results}
\,\,\,\,\,\, \textbf{Metrics.} Following prior work~\cite{li2024focaldreamer, sella2023vox}, we adopt CLIP-based metrics for quantitative evaluation:  \textbf{CLIP\textsubscript{sim}}~\cite{radford2021clip}, which measures the semantic alignment between a rendered view of the edited mesh and the target text prompt; and \textbf{CLIP\textsubscript{dir}}, which quantifies editing effectiveness by computing the directional CLIP similarity~\cite{gal2021stylegannada} between the original and edited meshes with respect to their text descriptions. In addition, we report \textbf{NIQE}~\cite{mittal2013niqe} and \textbf{NIMA}~\cite{talebi2018nima}, two no-reference image quality metrics that assess perceptual fidelity and better align with human visual judgment.

\textbf{Results.} Tab.~\ref{tab:compare_quant} summarizes the quantitative comparisons. Our method consistently achieves the best performance across all metrics, highlighting its superior ability to perform edits that are both semantically faithful to the target objectives and aesthetically satisfying to human vision, meeting the standards for high fidelity.

\subsection{Qualitative Results} 
\label{subsec:quant_results}

\cref{fig:compare} presents a qualitative comparison with baseline methods. As illustrated, the baselines struggle with complex tasks, resulting in coarse geometry and a lack of detail. The generated colors are often overly simple, flat, and visually inconsistent. In contrast, our method produces harmonious global structure, rich local details, and high-fidelity colors. 

For the fourth row, where the task is to delete the volcano, MagicClay~\cite{barda2024magicclay} replaces the volcano with a distorted rock; Instant3dit~\cite{barda2025instant3dit} substitutes the volcano with a patch of grass but alters parts of the original mesh. In contrast, our method seamlessly removes the volcano, filling the space with rocks similar to those in adjacent regions, achieving both visual and geometric harmony.

\subsection{Drag-based Mesh Editing}
Beyond text-driven editing, our framework can be naturally extended to support more sophisticated mesh manipulation tasks. To demonstrate its flexibility and broader applicability, we apply our approach to enable \textbf{drag-based mesh editing} through \textbf{drag-based image editing}.

Unlike prompt-based editing, drag-based image editing allows users to specify edits by drawing arrows to encode the desired spatial deformations, offering precise and intuitive control. In our implementation, we use LightningDrag~\cite{shi2024lightningdrag} as the underlying drag-based image editor. The workflow of drag-based mesh editing is as follows. First, drag-based image editing is performed to generate the desired image modifications. Second, mesh deletion is applied to corresponding regions of the original mesh. Finally, mesh insertion is then performed using the edited region meshes generated from drag-based image editing.

\cref{fig:drag}a shows the original meshes with arrow annotations illustrating the intent to open the angel’s wings and lift the cat’s paws, while \cref{fig:drag}b presents the results of drag-based editing. The successful results highlight the adaptability of our framework, demonstrating its strong potential for generalization to other advanced mesh editing scenarios.

\begin{figure}[t]
    \centering
    \includegraphics[width=\widthdrag\textwidth]{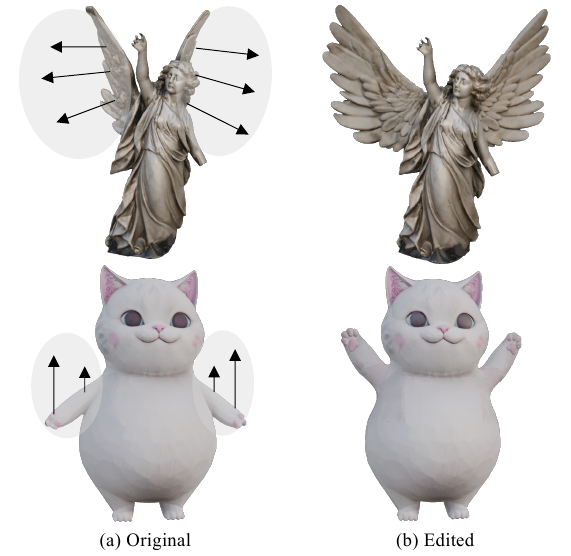}
    \caption{\label{fig:drag} Our method effectively adapts to drag-based mesh editing, enabling precise and intuitive geometry manipulation through user-specified drag controls. (a) shows the original mesh, with arrows drawn to signify drag; (b) shows the editing results. }
    \vspace{-1em}
\end{figure}

\subsection{Ablation Studies}
We conduct ablation studies on key components of our method: {Poisson Geometry Blending}, {Poisson Texture Harmonization} and the Mesh Generation Backend. Qualitative comparisons are presented in \cref{fig:abl}, and quantitative results are shown in Tab. \ref{tab:compare_abl}. To ensure fairness, we establish the baseline by applying mesh Boolean~\cite{cherchi2022interactive} between the original mesh and the edited region mesh.

\textbf{Effects of Poisson Geometry Blending.}
This module effectively resolves abrupt geometric discontinuities (\cref{fig:abl}), replacing them with realistic details such as cloth wrinkles, achieving a seamless transition. Compared with the baseline, it yields substantially improved geometric fidelity and quantitative performance.

\textbf{Effects of Poisson Texture Harmonization.}
This module improves texture harmony, for example, harmonizing the bright tone of the hand with the body’s darker gray while achieving texture continuity along boundaries. Further ablation experiments on each loss term demonstrate their significance: removing the distribution loss (\cref{fig:abl}e) produces unharmonious hues, omitting the gradient loss (\cref{fig:abl}f) results in blurred colors, and excluding the boundary loss (\cref{fig:abl}g) introduces visible texture discontinuities.

\textbf{Effects of Mesh Generation Backend.}
To demonstrate that our improvements stem from the proposed fusion framework rather than a specific generative backbone, we evaluate three distinct backends: Trellis~\cite{xiang2025structured}, Hunyuan3D 2.5~\cite{lai2025hunyuan3d} and MeshyAI~\cite{meshy2025}. As shown in Table~\ref{tab:compare_abl}, different configurations achieve consistently comparable state-of-the-art performance, thereby confirming that the observed improvements arise from our proposed fusion pipeline.
\begin{figure}[t]
    \centering
    \includegraphics[width=\widthabl\textwidth]{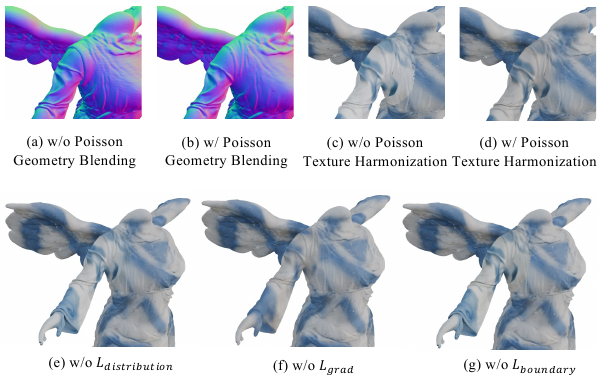}
    \caption{\label{fig:abl} Ablation studies. (a,b) Poisson Geometry Blending; (c,d) Poisson Texture Harmonization; (e) the distribution loss;  (f) the gradient loss; (g) the boundary loss.}
\end{figure}

\begin{table}[t]
\centering
\resizebox{\columnwidth}{!}{
\begin{tabular}{l|cccc}
\toprule
Method & CLIP\textsubscript{sim} $\uparrow$& CLIP\textsubscript{dir} $\uparrow$& NIQE $\downarrow$& NIMA$\uparrow$\\
\midrule
Baseline& 17.723 & 10.348 & 5.802 & 5.073 \\
w/ Geometry Blending & 20.502 & 11.979 & 5.774 & 5.290 \\
w/ Texture Harmonization & 19.399 & 10.724 & 5.290 & 5.184 \\
Ours (Hunyuan3D 2.5) & 19.903 & 18.622 & 6.108 & 5.749 \\
Ours (Trellis) & 19.166 & \textbf{18.911} & 6.246 & \textbf{5.989} \\
Ours (MeshyAI) & \textbf{20.801} & 18.479 & \textbf{4.710} & 5.928 \\

\bottomrule
\end{tabular}
}
\caption{Quantitative results of the ablation study. Both Geometry Blending and Texture Harmonization individually improve over the baseline, while combining them yields the best performance.\vspace{-1em}}
\label{tab:compare_abl}
\end{table}
\section{Conclusion}

We present CraftMesh, a framework for high-fidelity 3D mesh manipulation. Our approach combines 2D image editing with 3D generative models through a Poisson Seamless Fusion strategy. Built upon the hybrid SDF/Mesh representation, our method ensures both geometric and texture consistency, supporting complex and detailed edits that are seamlessly integrated into the original mesh. Experimental results demonstrate that CraftMesh outperforms existing baselines, delivering more coherent geometric structure, finer surface details, and higher-fidelity texture. The framework is also versatile, supporting advanced operations such as drag-based mesh editing. Future work may explore more sophisticated mesh editing techniques. 

\textbf{Limitation.} CraftMesh relies on off-the-shelf 2D and 3D generative models and therefore naturally inherits their limitations. For more complex meshes (e.g., those containing open, multiple layers, or noisy surfaces), this dependency may limit the achievable fidelity. Nevertheless, more powerful generative models can be readily substituted as they become available.

\section*{Acknowledgement} 

This work was supported by the National Natural Science Foundation of China (U25A20444, 62025207) and the Fundamental and Interdisciplinary Disciplines Breakthrough Plan of the Ministry of Education of China (JYB2025XDXM113).
    
{
    \small
    \bibliographystyle{ieeenat_fullname}
    \bibliography{main}
}


\end{document}